# Towards an Evidence-Based Approach to Climate Policy:
## Expanding the Scope of Evidence

Niklas Gaertner
Université Grenoble Alpes
niklas.gaertner@univ-grenoble-alpes.fr

Evidence-based policy (EBP) is an approach to policy decision-making that relies on evidence to make decisions. For example, during a pandemic, a politician adopting this approach might implement mask mandates based on studies demonstrating that masks reduce the transmission of viruses. When this practice is applied to climate change, where policymakers face decisions about management of climate impacts, mitigation, and adaptation, it is referred to as Evidence-based Climate Policy (EBCP). While the fundamental idea of EBCP is appealing and widely accepted, there is little discussion about what constitutes evidence in the context of policymaking, particularly given the variety of heterogeneous climate information. This is where I aim to contribute in this talk.

Historically, the principles of EBP originated from Evidence-Based Medicine (EBM). The core idea of EBM is to base clinical decisions on the best available scientific evidence. EBM integrates evidence-based guidelines, shifting decision-making from reliance on expert opinion to the use of scientific methods and statistical analysis (Sur & Dahm, 2011). Its success in medicine eventually led to the adoption of evidence-based approaches in other fields, including education, management, and policymaking.

My assumption is that due to this historical development of EBP the conception of evidence is rarely discussed in the literature. In medicine, evidence is clearly defined as empirical data derived from "formal research or systematic investigations using any type of science or social science methods" (Rychetnik et al., 2002). This conception also includes a hierarchy of evidence, where research studies are classified based on the strength and reliability of their methodologies (Fletcher & Sackett, 1979; Oxford Centre for Evidence Based Medicine, 2011). In this hierarchy, meta-analyses and randomized controlled trials are typically placed at the top, indicating the highest level of quality, while expert opinions and case reports are placed at the bottom, reflecting lower quality. However, the literature on EBP tends to use this concept of evidence without clarifying its meaning or critically examining whether this understanding is appropriate in the sense that the same standard of quality applies to the evidence needed for policy decision-making. This lack of clarity can lead to inconsistencies in how evidence is interpreted and utilized in policy decision-making. For instance, Montusch (2017) as well as Honnecker (2023) define evidence in EBP as "scientific knowledge", while Marchionni and Reijula describe it as the "best available evidence […] which typically prioritizes meta-analysis and randomized controlled trials over other evidence-generating methods" (Marchionni & Reijula, 2019). Similarly, La Caze & Colyvan (2016) refer to "scientifically respectable evidence". Based on these definitions, it remains unclear whether findings from ethnographic studies, for example, are considered as scientific evidence. What is more certain, however, is that even if evidence from social sciences is considered valid, it is viewed as inferior to the gold standard of randomized controlled trials and systematic reviews.

I argue that it is not desirable for EBCP to simply adopt the conception of evidence and the traditional evidence hierarchy used in EBM. To understand why this is the case, it can be helpful to take a step back and consider what evidence is in its most basic form. In philosophy, there are two general assumptions about evidence: (1) evidence is a propositional claim that supports or justifies another proposition (learning $E$ makes it more likely that $H$), and (2) evidence does not exist in isolation; it always exists relationally. Specifically, $p$ can only be considered evidence in the context of its relationship to $H$, where $p$ serves as evidence for $H$ (in the context of EBCP, $H$ represents a decision rather than a scientific hypothesis). Consequently, for something to qualify as evidence, it must be relevant and appropriate to the context in which it is applied, as its constitution and quality depend on its suitability for addressing the specific problem at hand. The context and issues in medicine differ significantly from those in climate policy. In the context of medicine, the goal is to obtain high-certainty evidence for cause-and-effect relationships to improve health through interventions like medical treatments. In contrast, the context of climate policy demands evidence that aids in understanding complex relationships and providing projections based on different socioeconomic scenarios. The objective is to address systemic, longer-term issues such as mitigating climate change and adapting to its effects. For example, evidence used to make informed policy decisions on climate change adaptation might include (i) expert judgments, (ii) probabilities and means from model-based ensembles, and/or (iii) scenario analysis. In the context of EBM, (i) is viewed as subordinate to other of evidence, but this may be unjustified in the context of EBCP, as expert judgment plays an important role in climate science (see Majszak & Jebeile 2023). However, (ii) and (iii) would not typically be part of the decision-making process according to the hierarchy of evidence established by EBM, as they do not align with its framework of evidence even though they provide valuable evidence for climate policy decisions (see Parker 2010, Duinker & Greig 2007). Other forms of evidence that may play an essential role in climate policy decision-making but are not adequately captured by the narrow conception of evidence in EBM include reanalysis data, paleoclimate evidence, impact assessments, social and economic data, local and indigenous knowledge, and attribution studies (see e.g. Cherubini et al. 2016; Kettle et al. 2014). However, being excluded or low-ranked by the evidence conception of EBM does not mean that these forms of evidence are not important in the context of policy-making.

Incorporating such evidence increases the likelihood of making effective climate policy decisions (Jones and Patwardhan 2014), and it is this relevance to decision-making that affects the quality of evidence in this context.

In a nutshell, I claim that (1) the concept of evidence used in EBM cannot simply be transferred to EBCP and (2) even if we broaden the scope of what we consider evidence, it is insufficient without also adapting the hierarchy system of evidence for EBCP. This work is based on a case study of PNACC-3 (Third National Plan for Climate Change Adaptation), a strategic plan developed by France to anticipate and mitigate climate change impacts. PNACC-3 provides guidelines for national and regional adaptation measures to ensure resilience across various sectors. By analyzing the types of evidence used (or recommended) to inform local policymakers, this case study highlights that some essential forms of evidence for effective climate adaptation decisions are not captured by the narrow conception of evidence in EBM, while others are not appropriately represented within its evidence hierarchy.